\documentclass[prl,twocolumn]{revtex4}
\usepackage{dcolumn}
\usepackage{amssymb}

\begin{document}

\title{Pentaquarks $uudd\bar s$ with one color sextet diquark}

\author{Claude \surname{Semay}}
\email[E-mail: ]{claude.semay@umh.ac.be}
\author{Fabian \surname{Brau}}
\email[E-mail: ]{fabian.brau@umh.ac.be}
\affiliation{Groupe de Physique Nucl\'{e}aire Th\'{e}orique,
Universit\'{e} de Mons-Hainaut,
Acad\'{e}mie universitaire Wallonie-Bruxelles,
Place du Parc 20, B-7000 Mons, Belgium}
\author{Bernard \surname{Silvestre-Brac}}
\email[E-mail: ]{silvestre@lpsc.in2p3.fr}
\affiliation{Laboratoire de Physique Subatomique et de Cosmologie,
Avenue des Martyrs 53, F-38026 Grenoble-Cedex, France}
\date{\today}

\begin{abstract}
The masses of pentaquarks $uudd\bar s$ are calculated within the
framework of a semirelativistic effective QCD Hamiltonian, using a
diquark picture. This approximation allows a correct treatment of the
confinement, assumed here to be similar to a Y-junction. With only color
antitriplet diquarks, the mass of the pentaquark candidate $\Theta$ with
positive parity is found around 2.2~GeV. It is shown that, if a color
sextet diquark is present, the lowest $uudd\bar s$ pentaquark is
characterized by a much smaller mass with a negative parity. A mass
below 1.7~GeV is computed, if the masses of the color antitriplet and
color sextet diquarks are taken similar.
\end{abstract}

\pacs{12.39.Pn,12.39.Ki,14.20.-c}


\maketitle

\section{Introduction}
\label{sec:intro}

Recent experiments have reported the existence of a very narrow peak in
$K^+n$ and $K^0p$ invariant mass distributions at 1.540~GeV
\cite{naka03}. This $\Theta$ resonance with isospin $I=0$, which has
been confirmed by
several experimental groups in various reaction channels \cite{exp},
is interpreted as a $uudd\bar s$ pentaquark \cite{diak97}. Quantum
numbers are not known, and various theories predict a $J^P=1/2^+$ or
$1/2^-$ assignment.

A popular model to describe the pentaquarks relies on the hypothesis
that quarks can form diquark clusters inside the pentaquark. In this
case, the confinement, which is at the origin a complicated five-body
interaction with 7 strings, reduces to a three-body interaction with 3
strings. A realistic confinement potential can then be built. The
diquark picture has been proposed to explain the properties of the
$\Theta$ resonance (let us note that this picture is not used in all
models \cite{stan04}). In Ref.~\cite{jaff03,shur04,koch04}, a good value
is obtained for the mass, but these models do not take into account the
full dynamics. In Refs.~\cite{naro04,naro04c,semaepja}, the confinement
is correctly taken into account and, as a consequence, pentaquark masses
are found above 2~GeV.

Such high masses are found because the pentaquark is assumed to be
composed of one antiquark and two identical color antitriplet $[ud]$
diquarks with vanishing spin and isospin. In that case, the two
identical diquarks must be in a relative $P$-wave in order to fulfill
the Pauli principle, and the lowest state has a total positive parity.
On the contrary, if two different diquarks are contained in a
pentaquark, this $P$-wave penalty can be avoided, but the lowest state
has a total negative parity.

In this work we show that pentaquarks can be composed with two different
diquarks, one in a color antitriplet representation and the other in a
color sextet representation. Provided some reasonable assumptions are
made about the diquark masses, a $uudd\bar s$ pentaquark with a negative
parity can be computed at mass near the $\Theta$ mass.

\section{$[ud]$ diquarks}
\label{sec:c6d}

All short-range interactions available between quarks, one-gluon
exchange \cite{flec88}, Goldstone-boson exchange \cite{gloz98}, and
instanton induced \cite{koch85} interactions, predict that the most
probable diquark which can be formed is the $[ud]$ pair in a color
$\bar 3$ representation with vanishing spin and isospin. Such a
structure with those quantum numbers is allowed by the Pauli principle
and is denoted $D$.

The instanton interaction between two quarks is a zero range interaction
given in Ref.~\cite{bein96}. It has the structure of
projectors on flavor-spin-color, and depends on two dimensioned
constants $g$ and $g'$. The spatial dependence of the potential is
singular and is generally regularized by a Gaussian
function $\rho(r)$ \cite{bein96}.

\begin{table}[htb]
\begin{ruledtabular}
\caption{Possible $[ud]$ diquarks, with their
quantum numbers (color representation $C$, isospin $I$ and spin $S$).
The contributions of the instanton induced force (Inst.)
and of the one-gluon exchange (OGE) potential are indicated.}
\label{tab:dq}
\begin{tabular}{lllllll}
Notation &  $C$ & $I$ & $S$ & Inst. & OGE \\
\hline
$D$ & $\bar 3$ & 0 & 0 & $-4\, g\, \rho(r)$ &
$-2/3\, \alpha_S/r$ \\
$D'$ & $6$ & 0 & 1 & $-2\, g\, \rho(r)$ &
$+1/3\, \alpha_S/r$ \\
\end{tabular}
\end{ruledtabular}
\end{table}

The instanton induced interaction is also attractive for a $[ud]$ pair
in color
$6$ representation with spin 1 and vanishing isospin. Such a structure
denoted $D'$ is also allowed by the Pauli principle and could exist
\cite{hong80,grom88}. Despite a
repulsive Coulomb contribution (see Table~\ref{tab:dq}), a diquark with
these quantum numbers could also be formed into an exotic hadron. A very
simple model developed below shows that color
antitriplet and color sextet diquarks could have similar masses.

\section{One-gluon exchange potential}
\label{sec:oge}

The total contribution of the one-gluon exchange process, at zero order,
is
\begin{equation}
\label{voge}
V_{{\rm OGE}} = \alpha_S \sum_{i<j=1}^3 \frac{\tilde \lambda_i \tilde
\lambda_j}{4} \frac{1}{|\vec r_i-\vec r_j|},
\end{equation}
where $\alpha_s$ is the strong coupling constant and $\tilde \lambda_i$
a color operator for the $i$th colored object.
In a $DD\bar s$ pentaquark, the contributions of the color operators
are the same for the three possible pairs, like in a (anti)baryon:
$\langle \tilde \lambda_i \tilde \lambda_j/4 \rangle = -2/3$. In a
$DD'\bar s$ pentaquark, the situation is different:
$\langle \tilde \lambda_{D} \tilde \lambda_{\bar s}/4 \rangle =+1/3$,
$\langle \tilde \lambda_{D'} \tilde \lambda_{\bar s}/4\rangle =
\langle \tilde \lambda_{D} \tilde \lambda_{D'}/4\rangle =-5/3$. The sum
of the color factors for the three pairs is $-2$ for a $DD\bar s$
pentaquark and $-3$ for a $DD'\bar s$ pentaquark. So we can expect, at
first approximation, a stronger attraction for the last system. This
Coulomb attraction is reinforced by the possibility of a $L=0$ total
angular orbital momentum for the $DD'\bar s$ system.

\section{Confinement}
\label{sec:conf}

The probably best possible simulation of the confinement in a three
colored object system is a Y-junction potential. Each colored source
generates a flux tube and the three strings connect at some point with
position $\vec r_0$, in order to minimize
the potential energy. It is expected that the energy density of the tube
is proportional to the color Casimir operator of the source
$\tilde \lambda_i^2/4$ and the length of the flux tube
\cite{bali00,semaca}.
The Y-junction potential, at zero
order, is then written
\begin{equation}
V_{{\rm Y}} = \frac{3\, a}{4} \min_{\vec r_0} \left[ \sum_{i=1}^3
\frac{\tilde \lambda_i^2}{4} |\vec r_i-\vec r_0| \right],
\label{vy}
\end{equation}
where $a$ is the usual string tension.

This complicated three-body interaction can be simulated by another
type of junction in which the three flux tubes connect to the center of
mass of the system. This one-body approximation is quite good (better
than 2\%)
provided the string tension is slightly renormalized \cite{silv04}. So,
the confinement we use in this work is
\begin{equation}
V_{{\rm CM}} = f \frac{3\, a}{4} \sum_{i=1}^3 \frac{\tilde
\lambda_i^2}{4} |\vec r_i-\vec r_{{\rm CM}}|,
\label{vcm}
\end{equation}
where $\vec r_{{\rm CM}}$ is center of mass coordinate. The parameter
$f<1$ rescales the interaction to simulate at best the Y-junction. The
possible values for the color Casimir operator are
$\langle \bar s| \tilde \lambda^2/4 | \bar s \rangle =
\langle D| \tilde \lambda^2/4 | D\rangle = 4/3$
and
$\langle D'| \tilde \lambda^2/4 | D'\rangle = 10/3$. With these numbers,
an increase of the confinement interaction can be expected for a
$DD'\bar s$ pentaquark with respect to a $DD\bar s$ pentaquark.

\section{Residual interactions}
\label{sec:resint}

The instanton interactions, which are assumed to be responsible for the
existence of the diquarks, must act also between a quark inside a
diquark and the other quarks or the antiquark of the pentaquark. In
Ref.~\cite{semaepja}, the residual instanton interactions are computed
for the lowest $DD\bar s$ pentaquark. They act between the antiquark and
any of the $u$ and $d$ quarks inside the diquarks (no such contribution
is expected between the two diquarks since they are in a $P$-wave). This
decreases the mass of such a state by a quantity which is around 40~MeV.
In the lowest $DD'\bar s$ pentaquark, the attractive instanton
interactions act between all the component quarks and antiquark since
the $DD'$ system is in a $S$-wave. A first crude estimation of this
effect, which depend on the total spin of the pentaquark, gives a total
potential whose strength is 2-3 times the corresponding one in a
$DD\bar s$ pentaquark. So the mass of the lowest $DD'\bar s$ system
could be decreased by a quantity which is around 100-150~MeV.

Supplementary residual interactions stemming from the one-gluon exchange
process can act in pentaquarks. A gluon carrying both spin and color can
turn a $D$ diquark into a $D'$ diquark. This mechanism can couple
$DD\bar s$ and $DD'\bar s$ pentaquarks and can decrease the mass of the
lowest system, that is to say the $DD'\bar s$ as we shall see below. It
can also produce an oscillation $D \leftrightarrow D'$ inside a
$DD'\bar s$ pentaquark. Such a mechanism can be generated by an
color-spin interaction of type
$\tilde \lambda_i \tilde \lambda_j \, \vec s_i \vec s_j$ between the
quarks inside the diquarks. As this potential is a relativistic
correction to the one-gluon exchange potential, its contribution to the
mass is expected weaker than those due to the interactions considered
above.

All these contributions are not easy to calculate,
but they could lead to the existence of several pentaquarks with masses
differing only from several tens of MeV. This must be confirmed
experimentally.

\section{Total Hamiltonian}
\label{sec:toth}

We use an effective QCD Hamiltonian derived in Ref.~\cite{naro04}, but
with all its auxiliary fields eliminated, as defined in
Ref.~\cite{sema04}.
The total Hamiltonian, in the center of mass frame, is a
semirelativistic kinetic energy part
supplemented by the Coulomb part of the one-gluon exchange potential
$V_{{\rm OGE}}$ and the
confinement $V_{{\rm CM}}$ described above
\begin{equation}
\label{h0}
H_0 = \sum_{i=1}^3 \sqrt{\vec p_i^{\,2}+ m_i^2} + V_{{\rm OGE}} +
V_{{\rm CM}}.
\end{equation}
Due to their probable weak contributions, the complicated residual
interactions discussed above are not taken into account here.

The particle self-energy is computed and appears as a
contribution depending on the constituent particle mass
\begin{equation}
M = M_0 +
\sum_{i=1}^3 \frac{C(s_i,m_i,a,\delta)}{\langle \sqrt{\vec
p_i^{\,2}+ m_i^2} \,\rangle}, \label{m0}
\end{equation}
where $M_0$ is an eigenvalue of $H_0$ and where
$C(s_i,m_i,a,\delta)$ is a negative contribution for a
fermion and vanishes for a boson \cite{naro04}. The
inverse gluonic
correlation length $\delta$ is around 1~GeV (the results are not
sensitive to this parameter). In pentaquarks with diquark clustering,
the only contribution of the self-energy comes from the antiquark. This
Hamiltonian was used in Ref.~\cite{semaepja} for the $DD\bar s$ systems.

\section{Parameters}
\label{sec:para}

The values of the parameters $a$, $\alpha_S$, $\delta$ and $m_n$ ($n$
stands for $u$ or $d$) are
taken from a previous work about pentaquarks \cite{naro04}. When the
three colored sources are identical, a good value for the parameter $f$
is around 0.94 \cite{silv04}. In this paper, we use this value for both
$DD\bar s$ and $DD'\bar s$ pentaquarks.

The parameters $m_D$ and $m_s$ are computed in order to
reproduce
the masses of the baryons $N$ and $\Lambda$ considered as
$Dq$ systems. The procedure to compute these baryon masses relies on the
operators~(\ref{h0}) and (\ref{m0}) adapted to a meson-like system, as
in Ref.~\cite{semaepja}. To consider the $N$ and $\Lambda$
baryons respectively as pure $Dn$ and $Ds$ states is
probably not the better approximation \cite{lich83}. But our aim is
just to obtain a reasonable estimation for the mass of the diquark $D$.
A mass of 350~MeV is in good agreement with
some other works \cite{shur04,naro04,semaepja,lich83}.
Table~\ref{tab:para} summarizes the parameters used in our calculations.


The color sextet diquark $D'$ does not appear in any baryon. So
it is not possible to fix its mass, as for the diquark $D$. Since a
simple model yields similar masses for the $D$ and $D'$ diquarks (see
below), $m_{D'}$ is considered here as a
free parameter, with a value around $m_D$.

\begin{table}[htb]
\begin{ruledtabular}
\caption{Parameters of the model.}
\label{tab:para}
\begin{tabular}{ll}
$a=0.15$ GeV$^2$ & $m_n=0$ \\
$f=0.94$         & $m_s=0.260$ GeV \\
$\alpha_S=0.39$  & $m_D=0.350$ GeV \\
$\delta= 1$ GeV  & $m_{D'} = m_D$ \\
\end{tabular}
\end{ruledtabular}
\end{table}

\section{Estimation of diquark masses}
\label{sec:compmdq}

An estimation of the diquark masses can be obtained by a simple
potential model, in the same spirit as in
Refs.~\cite{flec88,grom88,lich83}.
The two-quark Hamiltonian takes the following form
\begin{equation}
H_0 = \sum_{i=1}^2 \sqrt{\vec p_i^{\,2}+ m_i^2} + W_{{\rm Inst.}} +
W_{{\rm OGE}} + \epsilon_C \, a \ r. \label{h02} \\
\end{equation}
The mass operator takes also into account the quark self-energy (see
formula~(\ref{m0}) but for two particles).
The instanton contribution $W_{{\rm Inst.}}$ and the one-gluon exchange
contribution $W_{{\rm OGE}}$ are given in Table~\ref{tab:dq}. As
diquarks are not color singlet, it is not easy to determine the better
form for the ``confinement''. In Refs.~\cite{flec88,lich83}, a potential
$a\,r /2$ is introduced for the diquark $D$, based on a
$\tilde \lambda_i \tilde \lambda_j$ structure for color. But such a
prescription induces an anticonfining interaction for the $D'$, and
clearly
is not satisfactory. Thus, the situation concerning the confinement in
diquarks is quite unclear. Here we assume that the
confinement is given by $\epsilon_C \, a \ r$, in which $\epsilon_C$ is
a parameter in the range [0-1] which could depend on the total color
$C$ of the diquark \cite{grom88}.

Because this Hamiltonian takes into account the instanton forces, the
parameters chosen are taken from Ref.~\cite{semaepja}, except
$\epsilon_{\bar 3}=0.65$ and $\epsilon_6=0.40$ which are taken from
Ref.~\cite{grom88}. We then find $m_D=0.531$~GeV and
$m_{D'}=0.430$~GeV. This value of $m_D$ is larger that the one used in
this work, but the important result is that $m_{D'} < m_D$.
The diquarks $D$ and $D'$ have the same mass at 0.350~GeV with
$\epsilon_{\bar 3}=0.38$ and $\epsilon_6=0.33$. Let us
mention that, within this model,
the diquark $D$ ($D'$) can be a quite small object with a size
$\sqrt{\langle r^2 \rangle}$ varying from 1.0~fm (1.6~fm) to 0.5~fm
(0.8~fm) when $\epsilon_C$ increases from 0.2 to 1.

\section{Results}
\label{sec:res}

The numerical algorithm to solve the two-body problem is based on the
very accurate Lagrange-mesh method \cite{sema01}, and the technique to
solve the three-body problem relies on an expansion of the wave function
in terms of harmonic oscillator states with different sizes
\cite{nunb77}. After careful checks of the convergence properties, we
conclude that, with 20~quanta bases, a precision around 10~MeV can be
obtained for pentaquark masses.
The masses of some $uudd\bar s$ pentaquarks are presented in
Table~\ref{tab:mass}.

The lowest $DD\bar s$ is characterized by
$J^P=1/2^+$ and has a mass far above the experimental value 1.540~GeV.
This result is in agreement with those of
Refs.~\cite{naro04,naro04c,semaepja}.

The mass of the lowest $DD'\bar s$ is given with the condition
$m_D=m_{D'}$. This state is characterized by $J^P=1/2^-$. The low mass
obtained is due mainly to the fact that the $P$-wave penalty is avoided
(The $1/2^+$ state with $L=1$ is around 300~MeV above the $1/2^-$ state
with $L=0$). But other effects are necessary to explain this value. The
strong Coulomb potential plays an important role in the decreasing of
the masses. If we put arbitrarily
$\langle \tilde \lambda_i \tilde \lambda_j/4 \rangle = -2/3$, as in a
$DD\bar s$ pentaquark, the mass of the lowest $DD'\bar s$ pentaquark
increases by around 150~MeV. Moreover, contrary to what it is expected,
the interplay between the confinement, the Coulomb forces, and the
$S$-wave configuration makes the confining energy smaller with a
$DD'\bar s$ internal color configuration that with a $DD\bar s$ one. The
gain in binding energy is around 150~MeV.

\begin{table}[htb]
\begin{ruledtabular}
\caption{Masses in GeV (rounded to the nearest 10~MeV) for the lowest
$uudd\bar s$ pentaquarks, with
their spin-parity $J^P$ and quark content ($m_D=m_{D'}$).}
\label{tab:mass}
\begin{tabular}{lllllll}
Content & $J^P$ & Mass & & Content & $J^P$ & Mass \\
\hline
$DD\bar s$ & $1/2^+$ & 2.380 & & $DD'\bar s$ & $1/2^-$ & 1.680 \\
           & $1/2^-$ & 2.670 & &             & $1/2^+$ & 1.950 \\
\end{tabular}
\end{ruledtabular}
\end{table}

Obviously, the masses of the $DD'\bar s$ pentaquarks depend on the $D'$
mass. When $m_{D'}$ varies from 0.2 to 0.5~GeV, the mass of the $1/2^-$
$DD'\bar s$ varies from 1.550 to 1.800~GeV. The dependence is a
quasi-linear one, with a slope $\lesssim 1$. With reasonable values for
the $D'$ mass, the lowest $DD'\bar s$ state lies not too far from the
experimental $\Theta$ mass. As mentioned above, some gain of binding
energy could appear with the residual instanton induced interactions.

\section{Concluding remarks}
\label{sec:conclu}

It has been shown that dynamical models of the pentaquark $uudd\bar s$
within the diquark picture predict masses above 2~GeV, when the two
component diquarks (scalar, isospin singlet, color antitriplet) are
identical \cite{naro04,naro04c,semaepja}.

It is shown here that an effective QCD Hamiltonian predict that the
lowest $uudd\bar s$ pentaquark is a $1/2^-$ state with a mass close to
the $\Theta$ mass, provided this pentaquark contains two different
isospin singlet diquarks, one in a color antitriplet
configuration and the other in a color sextet one. This large shift in
mass is obtained independently of the value chosen for the sextet
diquark mass, although the precise value does obviously depend on it.
The low mass obtained is due mainly to the fact that the $P$-wave
penalty is avoided. But the particular color structures of the Coulomb
potential and of the confinement are necessary to explain this value.

Our $J^P=1/2^-$ assignment is in agreement with the QCD sum rules which
predict a negative parity for the $\Theta$ resonance \cite{math04}. The
lattice QCD calculations also predict that the parity of this state is
most likely negative \cite{csik03}.

It is possible that a lower mass, closer to the experimental values,
could be computed if the contribution of the residual interactions for
$DD'\bar s$ pentaquark are taken into account (such a work is in
progress). If this is the case, it will be interesting to study the
decay of this state, in order to understand the reasons of the small
width of the $\Theta$ resonance. This kind of calculation is complicated
and out of the scope of this paper.

\acknowledgments
C.~S. and F.~B. thank the FNRS for financial support.
This work was supported by the agreement CNRS/CGRI-FNRS.

\end{document}